# Stabilization of $\alpha$-UH$_3$ in U-Hf Hydrides: Structural, Magnetic, Thermodynamic, and Transport properties

Shanmukh Veera Venkata Uday Kumar Devanaboina[1], Oleksandra Koloskova[2], Silvie Mašková-Černá[1,2], and Ladislav Havela[1*]

[1]*Department of Condensed Matter Physics, Faculty of Mathematics and Physics, Charles University, Prague, Czech Republic,*
[2]*Institute of Physics (FZU), Czech Academy of Sciences, Prague, Czech Republic*

*\* Corresponding author: ladislav.havela@matfyz.cuni.cz*

# ABSTRACT

Hf substitution stabilizes the metastable body-centered cubic (*bcc*) $\alpha$-UH$_3$ phase in uranium hydrides, enabling systematic measurements of its magnetic, thermodynamic, and transport properties. (UH$_3$)$_{1-x}$Hf$_{x\setminus}$ hydrides ($x = 0.10, 0.15, 0.30, 0.40$) were obtained by hydrogenation of $\gamma$-*bcc* U$_{1-x}$Hf$_x$ precursor alloys. Powder X-ray diffraction shows a progressive suppression of $\beta$-UH$_3$ phase with increasing $x$, with $\alpha$-UH$_3$ domination at $x = 0.30$ and $\beta$-UH$_3$ nearly fully suppressed at $x = 0.40$. Magnetization measurements show ferromagnetic behavior for all compositions with Curie temperatures in the range $T_C \approx 178\text{-}185$ K and a maximum near $x = 0.15$; however, the spontaneous magnetization is strongly reduced with Hf content, decreasing from 1.0 $\mu_B$/U in pure $\beta$-UH$_3$ to 0.46 $\mu_B$/U for (UH$_3$)$_{0.60}$Hf$_{0.40}$. Specific-heat data show a broadened Curie anomaly in the $\alpha$-UH$_3$-rich hydride samples, consistent with a distribution of $T_C$ values arising from ferromagnetic inhomogeneities. Specific heat also reveals a monotonic decrease in the Sommerfeld coefficient $\gamma$ with increasing Hf concentration, reflecting a reduction in the electronic density of states at the Fermi level, especially in (UH$_3$)$_{0.60}$Hf$_{0.40}$. The resistivity of (UH$_3$)$_{0.60}$Hf$_{0.40}$ is very large (m$\Omega$ cm), exhibits a robust negative temperature coefficient over 2–300 K, and shows only weak magnetoresistance, placing transport in a strongly incoherent, disorder-dominated regime.

# I. INTRODUCTION

Uranium (U) holds a central position in the actinide series, widely studied for its role as a nuclear fuel element and its fundamental electronic properties. Its importance in nuclear energy has made it one of the most extensively studied elements in the series. The 5*f* electron shells play a unique role in the actinides, where their spatial extension and strong spin-orbit coupling give rise to a wide variety of electronic phenomena. The actinide series displays a clear shift in chemical behavior from lighter to heavier elements. The lighter actinides (Th–Pu) possess delocalized 5*f* states that contribute to electronic delocalization, leading to multiple oxidation states and resulting in a metallic bonding character reminiscent of late 3*d* transition metals. In contrast, the heavier actinides (Am onward) exhibit localized 5*f* states with weakly bonding character, analogous to the behavior of 4*f* electrons in rare-earth elements. Thus, Uranium sits near the crossover region between localized and itinerant behavior. As a consequence, U-based materials exhibit a complex magnetic and electronic behavior, including itinerant magnetism, strong electronic correlations, heavy fermions, and unconventional superconductivity.

Among uranium-based systems, uranium trihydride (UH$_3$) occupies a prominent position as the first material exhibiting ferromagnetism driven by 5*f* electrons. Its magnetic behavior provides important insight into the interplay between the crystal and electronic structures, giving rise to magnetic ordering in light actinides. In elemental uranium, the partially occupied 5*f* states participate in metal bonding, forming a relatively broad conduction band that also includes 6*d* and 7*s* states. Although the 5*f* states are at the Fermi level $E_F$, the density of states at the Fermi level, $N(E_F)$, remains low, yielding a non-magnetic ground state (Pauli paramagnet). The electronic structure of uranium compounds and alloys is modified by enhanced U-U spacing and hybridization with the ligand states. If the first tendency prevails and the Stoner criterion is fulfilled, enhanced $N(E_F)$ yields magnetic ordering. In U hydrides, the effect of volume expansion is accompanied by specific bonding conditions with 6*d*-1*s* hybridization as an important ingredient. Withdrawing the 5*f* states from the 5*f*-6*d* hybridization contributes to the ferromagnetism with relatively high $T_C \approx 170$ K at moderate U-U distances [1]. The Sommerfeld coefficient of the electronic specific heat increases significantly from ≈10 mJ/mol·K² in α-uranium to ≈30 mJ/mol·K² in UH$_3$.

Besides their fundamental importance, uranium hydrides also have practical importance in nuclear energy and storage. UH₃ meets several important criteria that make it an attractive option for hydrogen storage, which have been used so far, particularly for tritium. Uranium hydride has a very low equilibrium pressure (< 0.1 Pa) at room temperature, which means it effectively captures tritium, thereby eliminating its losses. The stored tritium can be rapidly released by heating to 400–450°C, generating a practical supply pressure exceeding atmospheric pressure ($10^5$ Pa). For the future, UH₃ continues to stand out as an alternative H(D, T) storage medium, especially in nuclear technologies [2]. It can also serve as nuclear fuel in light-water reactors (LWRs) [3].

Among various uranium hydrides, uranium trihydride (UH₃) exists in two structural modifications, $\alpha$-UH₃ and $\beta$-UH₃. The stable form, $\beta$-UH₃, adopts the Cr₃Si-type structure and crystallizes in a large cubic unit cell, as identified by Rundle via neutron diffraction in 1951 [4] [5]. It contains eight uranium atoms per unit cell with a lattice parameter of $a = 664$ pm, and two crystallographically distinct uranium sites (2$a$ and 6$c$), each coordinated by 12 hydrogen atoms. The hydrogen atoms occupy distorted tetrahedral sites, each positioned at a U–H distance of 232 pm from four surrounding U atoms. The shortest inter-U distance ($d_{U-U}$) found in $\beta$-UH₃ is 331 pm, which is inferior to the Hill limit (about 340 pm), assuming the minimum U-U distance yielding U magnetic moments. Several years later, $\alpha$-UH₃ was identified by Mulford et al. [6] as a metastable phase formed by hydrogenation of uranium metal at low temperatures. It transforms irreversibly into $\beta$-UH₃, testifying to the metastable character of $\alpha$-UH₃. It crystallizes in a body-centered cubic (*bcc*) lattice ($a = 416$ pm) with two formula units per unit cell, where uranium occupies a single Wyckoff position (6$c$). Each uranium atom is coordinated by 12 hydrogen atoms arranged in a distorted icosahedron, with a U–H bond distance of 232 pm, practically equal to that in $\beta$-UH₃ at ambient pressure [7]. The minimum $d_{U-U}$ in $\alpha$-UH₃ is 360 pm, which is larger than in $\beta$-UH₃, exceeding the Hill limit. The theoretical densities of $\alpha$-UH₃ and $\beta$-UH₃ are 11.11 g cm$^{-3}$ and 10.92 g cm$^{-3}$, respectively.

The ferromagnetic behavior of uranium hydride has been studied thoroughly in its stable $\beta$-UH₃ phase with $T_C \approx 170$ K, and a spontaneous magnetization of about 1 $\mu_B$/U. In contrast, $\alpha$-UH₃ has never been studied in its pure form, earlier studies [7] [8] indicate that it typically coexists with $\beta$-UH₃, and it remains unclear whether both phases are ferromagnets with identical $T_C$ or if the observed ferromagnetism arises from the $\beta$-UH₃ admixture and $\alpha$-UH₃ is non-magnetic. The issue was resolved by stabilizing *α*-UH₃ by alloying with Zr [9], which proved that the magnetism of both UH₃ variants has very similar fundamental characteristics as $T_C$ and size of ordered moments.

Alloying with selected transition metals in both UH₃ phases surprisingly enhances $T_C$ to about 15 at% of the transition element, an indirect effect attributed to variations in the H/U ratio. Above 15%, the H positions become less stable, and the H concentration tends to decrease [10]. The increase of $T_C$ reaches only 180 K for Zr or Ti, while it can exceed 200 K for 15% Mo [11]. Among the multiple elements attempted for alloying UH₃, there is no report on Hafnium (Hf). Here we bring original structure, magnetic, and thermodynamic data, which can be compared with analogous U-Zr hydrides.

The most common way of synthesis of alloyed U hydrides is to start from U-T alloys, prepared as stable or metastable within the γ-U structure type (*bcc*), by exposure to H₂ gas under elevated pressure. The formation of such alloys has not yet been investigated. Hence, our effort brings original data on the precursor alloys.

*α*-UH₃ can be regarded as a volume-expanded derivative of the high-temperature body-centered cubic (*bcc*) γ-U phase, with hydrogen atoms occupying interstitial sites and expanding the uranium lattice by approximately 60%. However, stabilizing the γ-U phase (stable only in the range 1048–1408 K) at room temperature is inherently challenging. The experimental approach begins with synthesizing metastable bcc γ-U alloys by doping with suitable transition metal elements to retain the γ-phase at room temperature and below, thereby avoiding the typical γ → β → α phase transformation sequence of uranium. The next step is to hydrogenate these stabilized alloys to form *α*-UH₃. Previous studies on γ-U alloys and their hydrides doped with selected d-elements (Mo, V, and Zr) [12] [13] [9] have shown that only Zr-based uranium hydrides successfully stabilize the α-UH₃ phase, whereas Mo- and V-doped systems typically yield the ferromagnetic *β*-UH₃ phase, each exhibiting different Curie temperatures. In this study, we explore the potential of hafnium (Hf) alloying with uranium (U) to stabilize the *bcc* *α*-UH₃ phase and investigate the effects of hydrogenation on the crystal lattice and the resulting changes in magnetic properties. Indeed, we found that Hf is similar to Zr in stabilizing the *α*-UH₃ phase and enhances $T_C$. We also focus on the behavior of magnetic susceptibility χ(T) in the immediate vicinity of $T_C$. We discuss effects related to inhomogeneities on the atomic scale.

## II. METHODS

Polycrystalline $U_{1-x}Hf_x$ alloys with Hf concentration $x = 0.10, 0.15, 0.30$, and $0.40$ were prepared by the standard arc melting technique using elemental metals (U: 99.8 wt.% purity, Hf: 99.9 wt.% purity) on a water-cooled copper hearth in an argon atmosphere. To ensure the homogeneity of the alloys, the sample buttons were turned over and re-melted three times. The weight loss during the melting is almost negligible. For the hydrogenation experiment, the alloy ingots were placed in an alumina crucible within a high-pressure, high-temperature reactor, which was initially evacuated to $10^{-6}$ mbar before being exposed to 120 bar of $H_2$. It was determined that the minimum $H_2$ pressure required for hydride formation, regardless of alloy composition, lies between 4-5 bars of $H_2$. Although higher pressures accelerate the process, the total hydrogen absorption remains unchanged. Opening the reactor after 96 hours revealed that the samples had disintegrated into a mixture of powder and small, brittle fragments, confirming extensive hydrogenation. To enable more precise monitoring of the hydrogenation process, including tracking pressure variations within a closed volume and ensuring its completion, we standardized on a 5-bar $H_2$ pressure in our experiments. Desorption experiments conducted by heating to 500 °C in a closed, evacuated system revealed that the hydrides absorbed approximately 3 H atoms per U atom, consistent with typical uranium hydride behavior, in which hydrogen is generally released around 450 °C. Thus, the hydrides are denoted as $(UH_3)_{1-x}Hf_x$.

The crystal structure and phase purity of the alloys were studied at room temperature using standard powder X-ray diffraction (XRD), with measurements taken directly from the surface of polished ingots, as the materials are very hard and cannot be crushed into a powder. The hydrides are obtained in the form of powder or small lamellae, so hydride samples were subsequently crushed and subjected to structural characterization by XRD. It is important to note that, unlike the fine powder of $β$-$UH_3$, the obtained hydrides are not pyrophoric, making them safe to handle and crush. The XRD studies of alloys and hydrides were performed using the Bruker D8 Advance diffractometer with Cu-K$_α$ radiation, with silicon as an internal standard (a = 5.4308 Å). XRD analysis confirmed that all $U_{1-x}Hf_x$ alloys exhibit the bcc $γ$-U phase as a primary phase, with a minor presence of the $α$-U phase, Hf metal, along with small amounts of surface impurities such as UC and $UO_2$. The crystal structure refinements were performed using the Rietveld method, implemented through the FullProf Suite software package [14]. The lattice parameters of the synthesized alloys and hydrides are consistent with values reported in the literature [U-Zr hydrides, other references] and are summarized in **TABLE** I.

Additional phase-purity analysis was performed by energy-dispersive X-ray spectroscopy (EDX) using a FEI Quanta 200 FEG scanning electron microscope (SEM). The sample surfaces were prepared by mechanical polishing followed by Ar-ion bombardment at an accelerating voltage of 4 kV. Magnetic properties of the material were subsequently analyzed using a SQUID Magnetometer [Quantum Design Magnetic Property Measurement System (MPMS) XL 7 T], with magnetization measurements performed across a temperature range of 2–300 K and in magnetic fields up to 7 T. The sample grains were fixed in random orientations using acetone-soluble glue. Specific heat measurements were carried out using a Quantum Design Physical Property Measurement System (PPMS 9T), with samples prepared as pellets from powdered hydride. One of the hydride samples, $(UH_3)_{0.60}Hf_{0.40}$, consisted of a lamella with a typical length of 2–3 mm, which enabled the attachment of electrical contacts using silver paste and subsequent measurements of the temperature-dependent electrical resistivity in a PPMS system.

## III. RESULTS AND DISCUSSION

### A. CRYSTAL STRUCTURE

**Fig**. 1 presents the room temperature XRD patterns of $(UH_3)_{1-x}Hf_x$ hydrides, demonstrating that all hydrides crystallize in a *bcc* structure, consistent with observations in Zr-doped hydrides [9]. The hydride prepared from pure U splat exhibits a mixture of $β$-$UH_3$ and $α$-$UH_3$ phases, with $β$-$UH_3$ being the dominant one. The substitution of Hf tends to suppress the $β$-$UH_3$ phase and promotes the $α$-$UH_3$ fraction. Both phases are present at x = 0.10, however, at higher Hf concentrations, $α$-UH3 predominates. For $x$ = 0.30, the XRD pattern displays only minor, broad peaks indicative of a small $β$-$UH_3$ admixture, suggesting that the sample is predominantly $α$-$UH_3$. In contrast, for $x$ = 0.40, the $β$-$UH_3$ phase is nearly completely suppressed. A detailed phase analysis indicates that a non-magnetic $HfH_2$ phase is consistently present as an impurity at approximately 33.1° in the diffraction patterns of nearly all hydrides. This phase is more abundant at x = 0.40 due to the higher Hf concentration in the hydride. A small peak observed at 55.40° could be attributed to the $HfH_2$ phase.

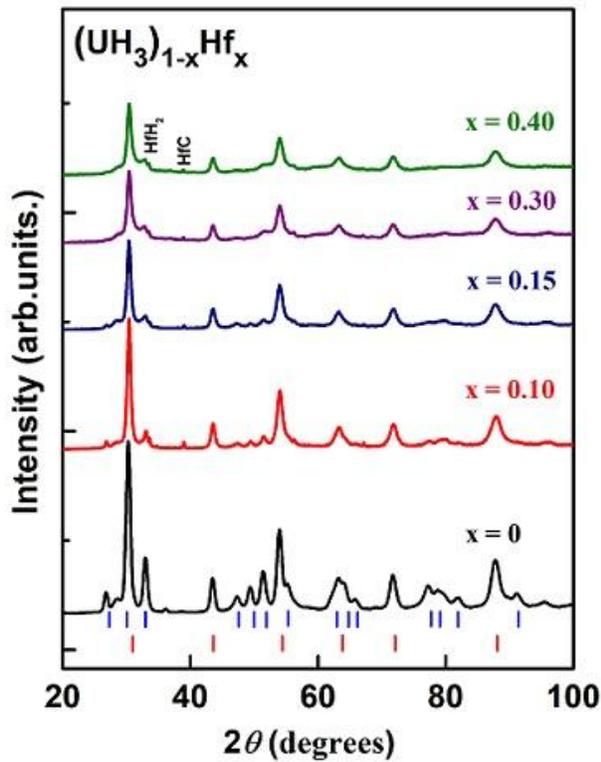

Fig. 1. X-ray diffraction patterns of the $(UH_3)_{1-x}Hf_x$ hydrides. The ticks indicate positions of diffraction lines expected for $\alpha$-$UH_3$ (red ticks - lower line) and $\beta$-$UH_3$ (blue ticks - upper line). Impurity peaks of $HfH_2$ and $HfC$ are indicated.

Furthermore, a minor amount of HfC as a surface impurity is detectable across all diffraction patterns. The lattice parameter $a \approx 416$ *pm* is consistent with reported values for $\alpha$-$UH_3$ (**TABLE** I). A slight increase in the lattice parameter is observed with higher Hf concentrations. Similar to U-Zr hydrides, it can be attributed to the larger atomic radius of Hf (159.0 pm) compared to U (156.0 pm). All $(UH_3)_{1-x}Hf_x$ hydrides exhibit relatively narrow diffraction peaks, contrasting with e.g., Mo-doped hydrides, where the peaks are much broader, but corresponding essentially to nanocrystalline $\beta$-$UH_3$ [12]. Analysis of peak broadening indicates that the mean crystallite size of all the hydrides falls within the nanoscale range, as detailed in **TABLE** I.

Irreversible changes in the phase composition of a pure uranium hydride sample were observed after approximately a few months, indicating that $\alpha$-$UH_3$ is unstable under ambient conditions. Additionally, the $UO_2$ peaks became more pronounced over time.

TABLE I: Room temperature lattice parameters $a$, mean crystallite size $D$ obtained from XRD analysis, Curie temperature $T_C$, spontaneous magnetization $M_s$, and paramagnetic Curie temperature $\Theta_p$ for the $(UH_3)_{1-x}Hf_x$ hydrides.

| Compound | $a$ (pm) Precursor | $a$ (pm) Hydride | $D$ (nm) | $T_C$ (K) | $M_s$ ($\mu_B$/U) | $\Theta_p$ (K) |
|---|---|---|---|---|---|---|
| $(UH_3)_{0.90}Hf_{0.10}$ | 348.4 | 415.7 | 22 | 178.5 | 0.91 | 186.1 |
| $(UH_3)_{0.85}Hf_{0.15}$ | 349.6 | 415.9 | 40 | 185.2 | 0.89 | 184.3 |
| $(UH_3)_{0.70}Hf_{0.30}$ | 351.2 | 416.6 | 35 | 183.7 | 0.75 | 183.3 |
| $(UH_3)_{0.60}Hf_{0.40}$ | 354.6 | 416.4 | 64 | 182.8 | 0.46 | 180.2 |

## B. MAGNETIZATION

Measurements of the temperature dependence of magnetization were performed on all synthesized uranium hydride samples to assess their magnetic behavior, as the parent U-Hf alloys are known to exhibit weak Pauli paramagnetism. The magnetization data were obtained using both zero-field-cooled (ZFC) and field-cooled (FC) protocols under a range of applied magnetic fields. **Fig**. 2 illustrates the temperature dependence of magnetization, $M(T)$, for $(UH_3)_{0.90}Hf_{0.10}$ and $(UH_3)_{0.60}Hf_{0.40}$ under different applied magnetic fields. The $(UH_3)_{0.90}Hf_{0.10}$ sample exhibits a mixed-phase composition containing both $\alpha$-$UH_3$ and $\beta$-$UH_3$, while the $(UH_3)_{0.60}Hf_{0.40}$ sample predominantly crystallizes in the $\alpha$-$UH_3$ phase. A pronounced bifurcation between zero-field-cooled (ZFC) and field-cooled (FC) magnetization curves is observed below the respective Curie temperatures in all investigated $(UH_3)_{1-x}Hf_x$ samples. This bifurcation persists down to the lowest measured temperatures, indicative of magnetic irreversibility commonly attributed to domain-wall pinning and substitutional disorder introduced by Hf. The small negative magnetization values observed in some ZFC curves are likely due to residual remanent fields present during nominal zero-field cooling in the SQUID magnetometer.

Low magnetic field ($\mu_0 H = 0.05$ T) magnetization data $M(T)$ are particularly convenient to reveal signatures of ferromagnetic ordering, with Curie temperatures ($T_C$) spanning from 178 K to 185 K (see **TABLE** I). Using the inflection point method of the $M(T)$ in low fields *(***Fig**. 2*)*, the highest $T_C$ of 185 K was observed in the $UH_3$ sample containing 15 at.% Hf, a mixed-phase composition of α-UH₃ with an admixture of β-UH₃. Measurements in high magnetic fields naturally show the ferromagnetic transition broadened. On the other hand, they yield more accurate magnetization data, free of demagnetization-factor effects, etc. In an applied field of 6 T, the 10% Hf-doped UH$_3$ sample exhibits a magnetization of 0.93 $\mu_B$/U; however, a systematic decrease is observed with increasing Hf concentration, reaching 0.46 $\mu_B$/U for the 40% Hf sample [See **Fig**. 2]. The reason for the decrease of U moment upon dilution of the U sublattice (we do not expect sizeable moments on Hf), there may be fundamental reasons, as a decrease of effective inter-site coupling affecting the moments in the itinerant magnetism case, seen e.g. in $U_{1-x}Th_xRhAl$ [15]. However, such a dramatic decrease was not observed in the Zr case [9], where the magnetization remained nearly unchanged across similar doping levels. Another specific reason may be a partial desorption of H in the measured sample, which reduces the total magnetization. (A fingerprint of a certain amount of a paramagnetic phase, a small magnetization decrease just at the 0 T point, can be seen in the hysteresis loop for 40% Hf in **Fig**. 5 below).

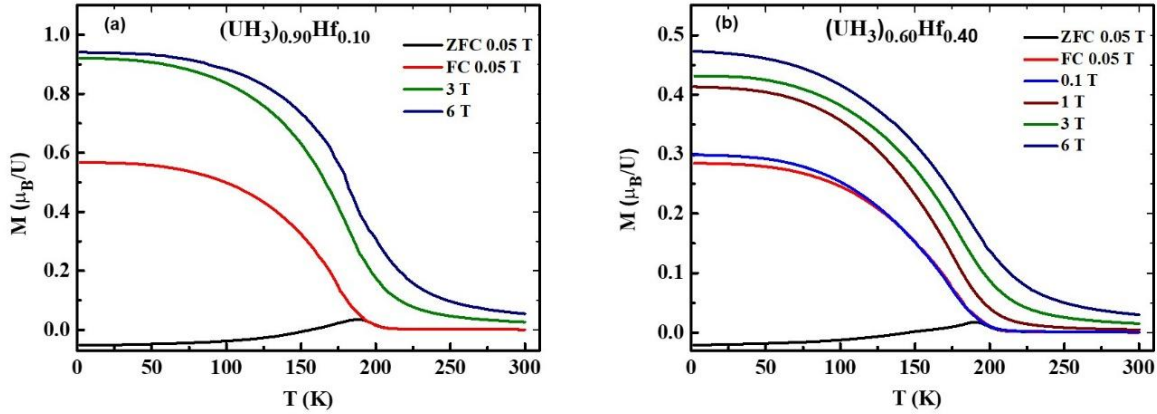

Fig. 2. Temperature dependence of magnetization, *M*(T), for the hydrides: (a) $(UH_3)_{0.90}Hf_{0.10}$ and (b) $(UH_3)_{0.60}Hf_{0.40}$, measured under magnetic fields of 0.05 T (in both ZFC and FC modes), as well as under various higher magnetic fields in FC mode. The plots clearly demonstrate a reduction in magnetization with increasing Hf concentration.

Furthermore, the magnetic susceptibility of all the U–Hf hydride samples was analyzed using the modified Curie–Weiss (MCW) law is given below:

$$\chi = \chi_0 + \frac{C}{(T-\theta_\mathrm{p})} \tag{1}$$

where $C$ (= $N_A \mu_{eff}^2 / 3k_B$) is the Curie constant, $N_A$ and $k_B$ are Avogadro's number and Boltzmann constant, respectively, $\theta_\mathrm{p}$ is the paramagnetic Curie temperature, and $\chi_0$ is a temperature-independent term, which typically accounts for contributions such as the Pauli paramagnetic term. This analysis was conducted to investigate the paramagnetic behavior of the samples above their respective Curie temperatures. Due to the stabilization of the $\alpha$-UH$_3$ phase in the (UH$_3$)$_{0.60}$Hf$_{0.40}$ hydride, this composition was selected as the primary focus for detailed investigation of its magnetic and transport properties. The inverse DC susceptibility, $\chi^{-1}(T)$, is investigated under various applied magnetic fields, revealing an intriguing behavior at low fields. Notably, a sharp downturn from Curie-Weiss (CW) behavior is observed well above the ferromagnetic Curie temperature, particularly in the low-field regime (0.01 T – 0.25 T). This anomaly gradually softens with increasing field strength (0.25 to 6 T), and $\chi^{-1}(T)$ progressively approaches a more typical CW-like linearity [see **Fig**. 3(a)]. Such field-dependent deviations have often been associated with the onset of Griffiths phase (GP)-like behavior [16] [17] and have also been observed in a variety of other systems [18] [19] [20] [21] [22]. The presence of finite-sized magnetic correlations above $T_C$ suggests the formation of short-range magnetic clusters embedded within the paramagnetic matrix. These clusters, while locally ordered, do not percolate to form a system-wide, long-range ferromagnetic state. As a result, no spontaneous magnetization ($M_{SP}$) emerges in this regime, where locally correlated regions exist without global order, which is a hallmark of Griffiths phase-like behavior. In the presence of magnetic inhomogeneities above $T_C$, arising from the formation of short-range correlated magnetic clusters, the system deviates from uniform paramagnetic behavior. These local fluctuations violate the assumption of spatially uncorrelated moments in the Curie-Weiss law, leading to nonlinearity in the inverse susceptibility, particularly at low magnetic fields. Consequently, fitting the low-field $\chi^{-1}(T)$ data yields unreliable estimates of critical parameters, such as the effective magnetic moment and the paramagnetic Curie temperature.

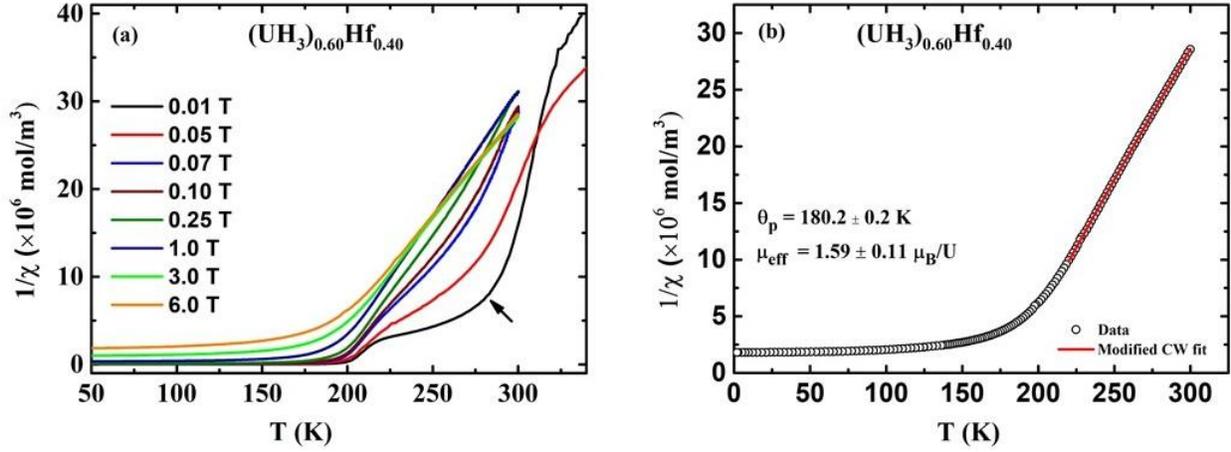

Fig. 3. (a) Inverse dc magnetic susceptibility ($1/\chi$) as a function of temperature for $(UH_3)_{0.60}Hf_{0.40}$ under various applied magnetic fields. The arrow indicates a downturn in the paramagnetic region, highlighting a deviation from ideal Curie-Weiss (CW) behavior that is suppressed at higher fields. (b) Modified Curie-Weiss fit of the inverse susceptibility data at higher magnetic field for the same hydride, yielding the effective magnetic moment per uranium atom $\mu_{eff}$ and the paramagnetic Curie-Weiss temperature $\theta_p$.

**Fig**. 3(b) shows the modified Curie-Weiss (MCW) fit of the high-field inverse susceptibility curve for the $(UH_3)_{0.60}Hf_{0.40}$ hydride, using the fitting range from 220 to 300 K to ensure a well-defined paramagnetic matrix. The resulting paramagnetic Curie temperature ($\theta_p$) is (180.2 ± 0.2) K, which is consistent with the literature data for the $\beta$-UH$_3$ or $\beta$-UD$_3$ (168 – 181 K) [23]. The positive $\theta_p$ indicates the dominance of ferromagnetic interactions in the paramagnetic region. The fitted $\theta_p$ values for the other hydride compositions are also positive and show a slight increase with decreasing Hf concentrations, as summarized in **TABLE** I. The effective paramagnetic moment normalized per uranium atom obtained from the Curie constant as the MCW fitting parameter is (1.59 ± 0.11) $\mu_B$/U. This value is lower than the theoretical range of 2.2–2.4 $\mu_B$/U [23], and the difference is compatible with a decrease in spontaneous magnetization, possibly due to H release.

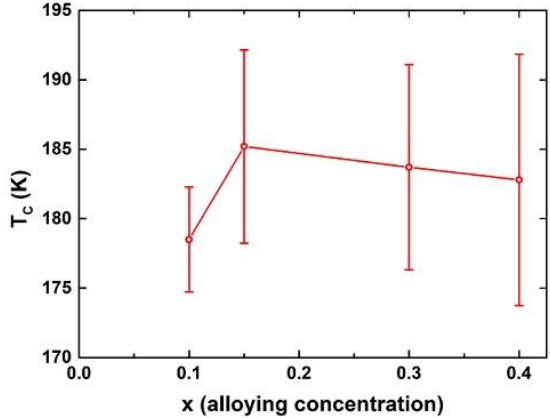

Fig. 4. Concentration dependence of the Curie temperature for the hydrides $(UH_3)_{1-x}Hf_x$

One of the primary focuses of this study is to investigate how the Curie temperature ($T_C$) varies with Hf concentration. For each Hf-alloyed hydride, $T_C$ was determined using various techniques - *the inflection point of the M(T) dependence in low magnetic fields (**Fig**.2), Arrott plots ($M^2$ against H/M around the expected $T_C$) (**Fig**. 5), AC-susceptibility, and the temperature dependence of specific heat ($C_p(T)$) (**Fig**.7a)*. In the inflection point method, $T_C$ values were determined as the average between the inflection point of the $M(T)$ curve at low magnetic fields and the intersection of the tangent (extrapolated from the highest slope) with the baseline. The technique of Arrott plots [24] is based on re-plotting $M(H)$ taken at a particular temperature in the $M^2$ vs. $H/M$ form, which should yield a straight line in higher fields. If the extrapolation of $H/M \rightarrow 0$ gives a positive $M^2$, $M$ is associated with spontaneous magnetization $M_s$ at the given temperature. Measurements were performed systematically over the temperature range 165–210 K in steps of $\Delta T = 3$ K (see **Fig**. 5). The temperature dependence of these intercepts was used to determine the Curie temperature, defined as the temperature at which $M_s = 0$. The results show increased uncertainty with higher Hf concentrations, as reflected by the error bars in **Fig**. 4. This behavior is consistent with observations in other transition metal-doped U-hydrides [9,12,13]. The maximum $T_C$ appears in the hydride of 15% Hf concentration and decreases further at higher concentrations.

At this point, we can only hypothesize why the maximum $T_C$ occurs at alloying element concentrations of 12–15%. The role of potential H vacancies in both the Hf/Zr and Mo cases remains uncertain. This is suggested by the formula we use, which indicates that the actual H concentration is more closely related to the U concentration than to the total concentration of metal atoms. Generally, one might attribute this phenomenon to variations in the mean U-U spacing, or shifts in the balance of conduction electrons, or a combination of both. The maximum occurs for both smaller and larger alloying elements, so the volume effect is unlikely.

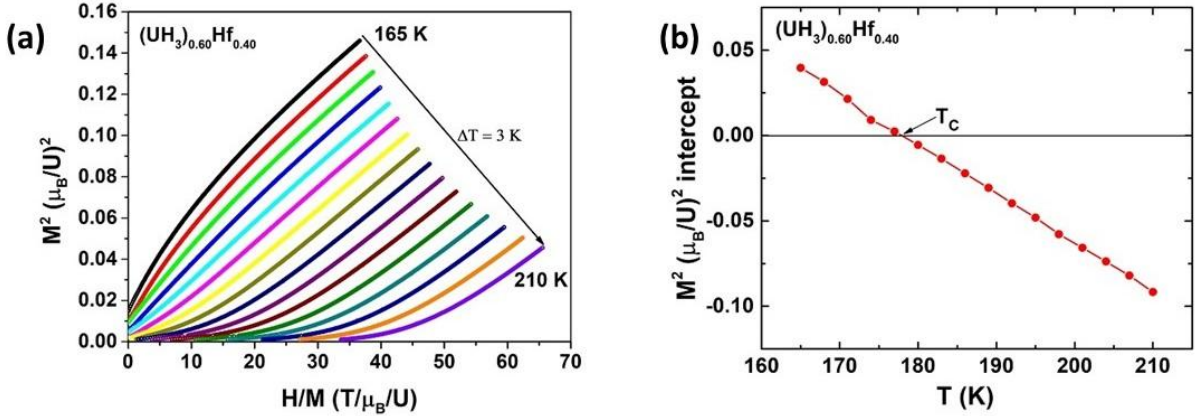

Fig. 5. (a) Arrott plots for the hydride $(UH_3)_{0.60}Hf_{0.40}$ derived from isothermal magnetization curves in the temperature range of 165 K to 210 K, with an increment of $\Delta T = 3$ K. (b) Temperature dependence of the $M^2$ intercepts obtained from the linear fitting of all the isotherms in (a).

The U-Hf hydrides exhibit very broad hysteresis loops at low temperatures. For example, **Fig**. 6 (a) and (b) show the hysteresis loops at various temperatures for $(UH_3)_{0.85}Hf_{0.15}$ and $(UH_3)_{0.60}Hf_{0.40}$. These loops do not exhibit true saturation at 7 T. One has to realize that we actually observe minority loops, and the actual width can be even larger, as shown for the $UH_3$-Zr system, measured in fields up to 13 T. Therefore, the magnetization obtained at low temperatures from M(H) in the FC mode exceeds the values recorded at the maximum fields of the hysteresis loops [9]. The erratic jumps observed in M(H) at 2 K are similar to those in the Zr-alloyed UH3, which are associated with difficult nucleation of remagnetization and trigger an avalanche-type process. This behavior is characteristic of materials with significant domain wall pinning. At higher fields, the magnetization increases more gradually, attributed to the rotation of magnetization towards the field direction.

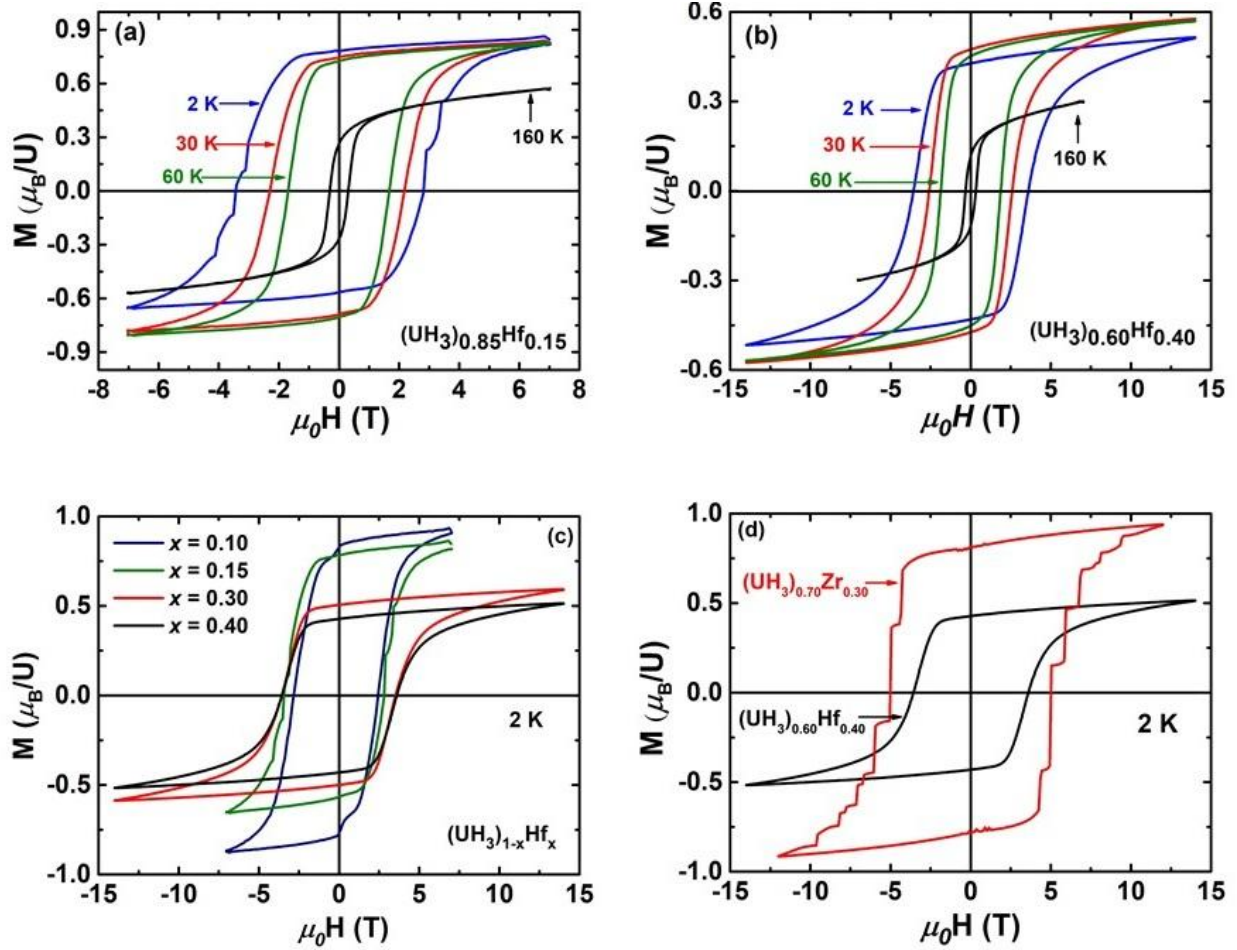

Fig. 6. Magnetic-field dependence of the magnetization $M$(H) (in $\mu_B$/U) for $(UH_3)_{1-x}Hf_x$ hydrides. (a) $M$(H) isotherms for $(UH_3)_{0.85}Hf_{0.15}$ at T = 2, 30, 60 and 160 K. (b) same for $(UH_3)_{0.60}Hf_{0.40}$. (c) $M$(H) loops at 2 K for all $(UH_3)_{1-x}Hf_x$ hydrides. (d) Comparison of the 2 K hysteresis loops for $(UH_3)_{0.60}Hf_{0.40}$ and $(UH_3)_{0.70}Zr_{0.30}$.

As expected, coercivity increases with decreasing temperature for all the hydride samples. The width of the loops also increases with increasing Hf concentration, reaching approximately 7.5 T. This value is notably lower than that observed in Zr-doped hydrides (see **Fig**. 6(d)), which can be attributed, as already mentioned, to lower maximum fields used, not sufficient to reach a full saturation.

## C. SPECIFIC HEAT

The specific heat at constant pressure, $C_p(T)$, was studied systematically for $(UH_3)_{1-x}Hf_x$ hydrides with $x = 0.15, 0.30, 0.40$ hydrides over the temperature range from 2 K – 300 K to investigate the Sommerfeld coefficient and its Hf concentration dependence from the low temperature electronic part, Debye's temperatures associated with the phonon term, ferromagnetic Curie temperatures and magnetic entropy from the magnetic contribution to the specific heat. **Fig**. 7 (a) shows the temperature variation of the specific heat of the three hydrides over a temperature range of 2 – 200 K, compared with $\beta$-$UH_3$ and the calculated $\alpha$-$UH_3$ curves for reference.

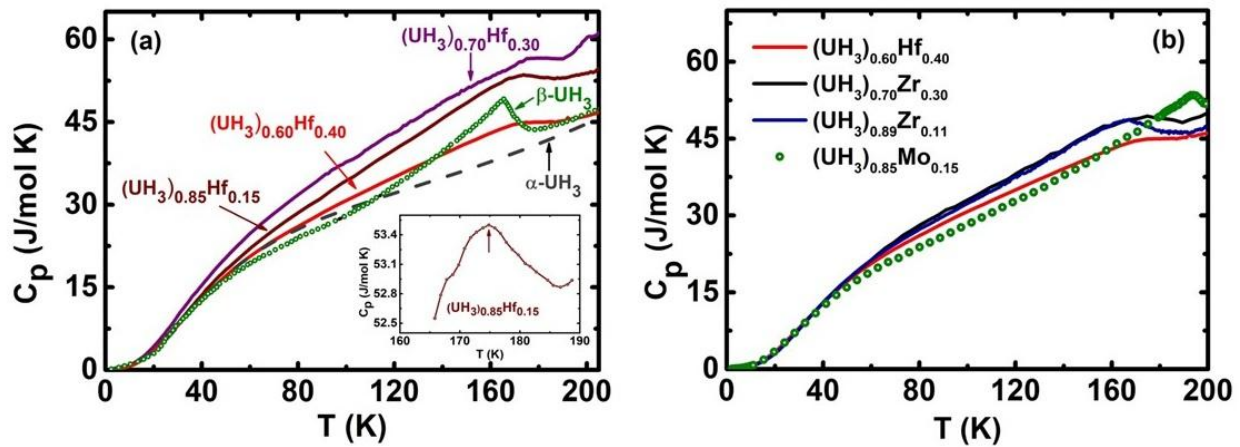

Fig. 7. (a) Temperature dependence of specific heat $C_p(T)$ for $(UH_3)_{1-x}$ $Hf_x$ hydrides with $x = 0.15$, 0.30, and 0.40, $\beta$-$UH_3$. For comparison, the green circles denote the literature data for $\beta$-$UH_3$ and the dashed lines denote the calculated model for $\alpha$-$UH_3$ as a non-magnetic analogue are also shown. Inset: enlarged view of $C_p(T)$ for $(UH_3)_{0.85}Hf_{0.15}$, highlighting a broad peak near its $T_C$. (b) Comparison of $C_p(T)$ for selected substituted compositions: $(UH_3)_{0.60}Hf_{0.40}$, $(UH_3)_{0.70}Zr_{0.30}$, $(UH_3)_{0.89}Zr_{0.11}$ and $(UH_3)_{0.85}Hf_{0.15}$

One can distinguish a broad $T_C$ anomaly around 180 K in $(UH_3)_{0.60}Hf_{0.40}$, which is different from the sharp cusp typically associated with $T_C$ in $\beta$-$UH_3$ [7]. In contrast, the $(UH_3)_{0.85}Hf_{0.15}$ hydride exhibits a peak near its $T_C$ [see inset in **Fig**. 7(a)] that more closely resembles the $\beta$-$UH_3$ trend, likely due to its lower Hf concentration and the significant presence of the $\beta$-$UH_3$ phase. The $\alpha$-$UH_3$-rich hydride $(UH_3)_{0.70}Zr_{0.30}$ shows a similarly broadened $T_C$ feature, comparable to that of $(UH_3)_{0.60}Hf_{0.40}$. All the $(UH_3)_{1-x}Hf_x$ hydrides show their $T_C$ slightly higher than that of $\beta$-$UH_3$. **Fig**. 7(b) compares $C_p(T)$ curves of the Zr and Mo-substituted hydrides with the $(UH_3)_{0.60}Hf_{0.40}$ hydride.

Despite the similar roles of Hf and Zr in stabilizing the pure α-UH$_3$ phase, the (UH$_3$)$_{0.89}$Zr$_{0.11}$ composition exhibits a well-defined cusp at $T_C$ near 170 K [9]. Whereas (UH$_3$)$_{1-x}$Hf$_x$ hydrides display broadened Curie anomalies and show other interesting Griffith's phase-like behavior above $T_C$ due to the ferromagnetic inhomogeneity related to U lattice dilution.

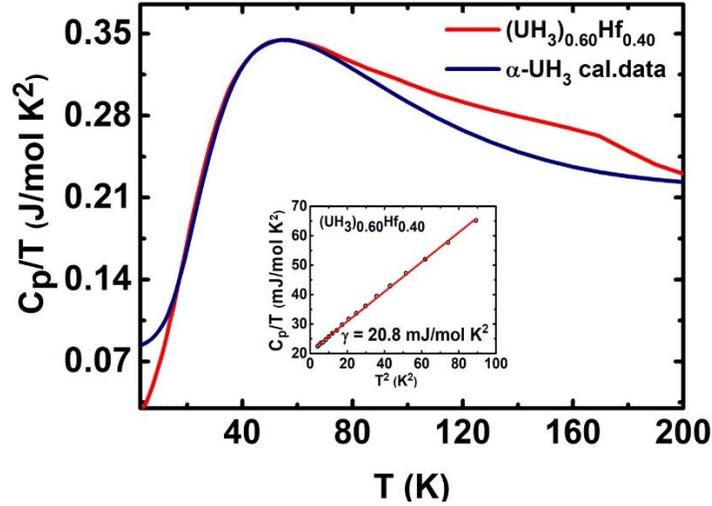

Fig. 8. Total specific heat of (UH$_3$)$_{0.60}$Hf$_{0.40}$ and the calculated model of α-UH$_3$ (as a non-magnetic analogue), plotted as $C_p/T$ vs. $T$. Inset: $T^2$ dependence of specific heat plotted as $C/T$ vs $T^2$, and includes the value of Sommerfeld coefficient γ

The calculated model contains only the electronic and phonon parts and does not contain a magnetic contribution, so it is very convenient to use it as a non-magnetic analogue. The low-temperature part of the specific heat can be approximated by the Debye model, from which it is possible to estimate the Sommerfeld coefficient (γ) of the electronic specific heat and Debye temperature ($θ_D$) as the phonon contribution. From the linear fit at a low temperature using the equation,

$$\frac{C}{T} = \gamma + \beta T^2 \qquad (2)$$

The values of γ and β = 1944/$θ_D^3$ J/mol K divided by the number of atoms per formula unit (f.u.), have been determined and are also listed in **TABLE** II. The inset of **Fig**. 7(c) shows the low-temperature linear fit of $C_p/T$ vs $T^2$ from which a Sommerfeld coefficient γ = 20.8 mJ/mol. K$^2$ per formula unit is obtained for (UH$_3$)$_{0.60}$Hf$_{0.40}$. This value is lower than those reported for β-UH$_3$ (29 mJ/mol U K$^2$) [25] [26], where the entire uranium atom is considered in the normalization. When normalized to the uranium contribution only, the $(UH_3)_{0.60}Hf_{0.40}$ yields the γ value of 34.5 mJ/mol U $K^2$. Since the Sommerfeld coefficient γ is proportional to the electronic density of states at the Fermi level, $n(E_F)$, the latter can be estimated from

$$n(E_F) = \frac{3\gamma}{\pi^2 k_B^2 N_A} \tag{3}$$

Where $k_B$ is Boltzmann's constant and $N_A$ is Avogadro's number. For $(UH_3)_{0.60}Hf_{0.40}$ hydride, the resulting density of states at the Fermi level is 8.8 states/ (eV/f.u.) with γ = 20.8 mJ/mol. $K^2$. The corresponding γ values for $(UH_3)_{0.70}Hf_{0.30}$ and $(UH_3)_{0.85}Hf_{0.15}$ are 24.3 mJ/mol $K^2$ and 26.3 mJ/mol $K^2$, respectively, obtained through the same fitting procedure. Taken together, the three compositions reveal a clear decrease in the γ value and a subsequent decrease in the density of states with increasing Hf substitution. This reduction is likely due to the dilution of the uranium sub-lattice by nonmagnetic hafnium atoms, which disrupts the 5f-electron network responsible for magnetic ordering. As a consequence, the magnetic exchange interactions are weakened, leading to suppressed magnetic ordering. This is consistent with the observed reduction in the magnetization for $(UH_3)_{0.60}Hf_{0.40}$, which decreases to 0.46 $\mu_B$/U compared to 1.0 $\mu_B$/U in the β-$UH_3$. The effective Debye temperatures for $(UH_3)_{1-x}Hf_x$ hydrides were obtained from the slopes (β) of the low temperature part of specific heat $C_p(T)$ fits using Eq. (2). In the calculation, only U and Hf ions are included, consistent with the treatment used for β-$UH_3$ [25] and for Zr and Mo-substituted analogues, while hydrogen ions are excluded because of the large ionic mass disparity as the low-energy acoustic branch is governed primarily by the heavy-atom (U/Hf) masses, whereas the light H atoms mainly contribute to high-frequency optical modes that are not appreciably populated at low temperatures.

TABLE II: Sommerfeld coefficient γ, Debye temperature $\theta_D$, and zero-field magnetic entropy $S_{mag}$ for $(UH_3)_{1-x}$ $Hf_x$ hydrides with $x$ = 0.15, 0.30, and 0.40. $S_{mag}$ is given per mole of formula unit in J/mol K and as a fraction R ln2.

| Compound | γ (mJ/mol $K^2$) | $\theta_D$ (K) | $S_{mag}$ (J/mol K) | $S_{mag}$ (Rln2) |
|---|---|---|---|---|
| $(UH_3)_{0.85}Hf_{0.15}$ | 26.3 | 163.6 | 4.23 | 0.71 |
| $(UH_3)_{0.70}Hf_{0.30}$ | 24.3 | 159.8 | 3.60 | 0.62 |
| $(UH_3)_{0.60}Hf_{0.40}$ | 20.8 | 156.5 | 3.10 | 0.51 |

Values of the $\theta_D$ fall in the range of 156-165K for $(UH_3)_{1-x}Hf_x$ hydrides (**TABLE** II). These values are slightly lower than those reported for $(UH_3)_{1-x}Zr_x$ analogues [9], consistent with the heavier Hf sub-lattice lowering the characteristic frequencies of the acoustic phonon modes. As the $C_p(T)$ was not measured for $U_{1-x}Hf_x$ alloys, comparison may be made with $U_{1-x}Zr_x$ alloys, which show Debye-like behavior with $\theta_D = 160–170$ K [9], close to those of $(UH_3)_{1-x}Hf_x$. This correspondence indicates that the low-lying acoustic phonon modes are essentially similar in both the precursor alloys and the hydrides. **TABLE** II summarizes the monotonic decrease of $\gamma$ and $S_{mag}$ with x, indicating a systematic suppression of the low-temperature electronic term and the recovered magnetic entropy across the series.

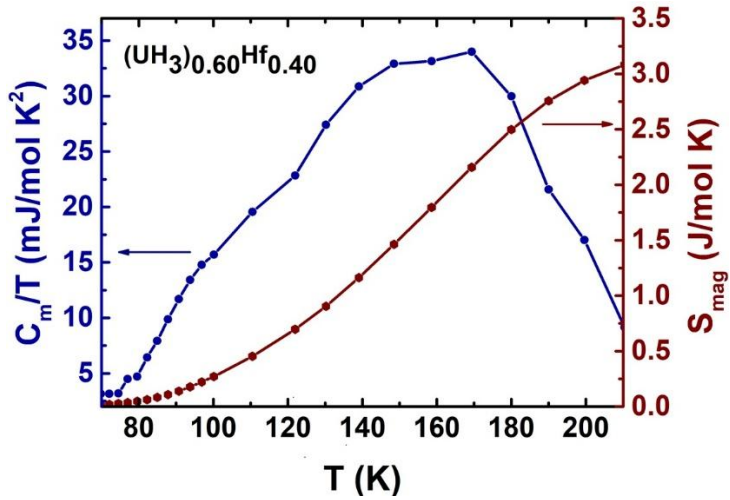

Fig. 9. Temperature dependence of magnetic specific heat $C_m/T$ (left axis) and the corresponding magnetic entropy $S_{mag}$ (right axis) for $(UH_3)_{0.60}Hf_{0.40}$ hydride.

To estimate the magnetic contribution to the specific heat of $(UH_3)_{0.60}Hf_{0.40}$, we compare the measured specific heat with a calculated model for $\alpha$-$UH_3$ that includes only lattice and electronic contributions and serves as a nonmagnetic reference. Subtracting this background yields the magnetic specific heat $C_m$. **Fig**. 9 shows $C_m/T$ and the corresponding magnetic entropy $S_{mag}$.

$$S_{mag}(T) = \int_0^T \frac{C_m(T')}{T'} dT' \qquad (4)$$

The recovered entropy is approximately 3.1 J/mol K, corresponding to ~ 0.51 R ln2. This value is significantly lower than the full entropy (R ln2) expected for a system with fully localized spin-½ magnetic moments. The reduced entropy suggests that the 5f electrons in $(UH_3)_{0.60}Hf_{0.40}$ exhibit itinerant or weakly localized behavior, consistent with band-like 5f magnetism. Moreover, the incomplete entropy recovery may be attributed to disorder-induced suppression of long-range magnetic order, possibly due to Hf-induced dilution of the uranium sub-lattice. This is also indicative of persistent short-range magnetic correlations or fluctuations above the ordering temperature, supporting the presence of a magnetically inhomogeneous or cluster-like state. These findings reinforce the picture of a partially ordered, diluted magnetic system with reduced 5f participation in the collective magnetic ground state.

**D. RESISTIVITY**

Temperature dependence of electrical resistivity $\rho(T)$ for $(UH_3)_{0.60}Hf_{0.40}$ hydride was measured over the range of 2-300 K using the standard DC four-probe technique in both zero-field and under applied magnetic fields, as shown in **Fig**. 10. A small lamella, approximately 2-3 mm long, was used for the experiment. The resistivity exhibits pronounced anomalous behavior as $\rho(T)$ decreases monotonically with increasing temperature over the entire measured range, yielding a negative temperature coefficient ($d\rho/dT < 0$) at all temperatures. The absolute values of $\rho$ are very large, reaching several mΩ-cm at low temperatures. No distinct anomaly or cusp attributable to magnetic ordering is resolved in $\rho(T)$. Application of a magnetic field results in only a weak positive magnetoresistance at low temperatures, while the negative temperature coefficient remains essentially unaffected. The magnitude of the resistivity itself provides an important constraint on the nature of charge transport. The low-temperature resistivity exceeds the Mott–Ioffe–Regel limit by more than an order of magnitude, indicating a strongly incoherent transport regime in which the electronic mean free path approaches interatomic dimensions and conventional Boltzmann transport concepts are no longer applicable [27] [28]. Such behavior places $(UH_3)_{0.60}Hf_{0.40}$ in the regime associated with the Mooij correlation, where very large residual resistivities are commonly accompanied by flattened or negative resistivity slope ($d\rho/dT < 0$) [29]. Such transport phenomenology is commonly observed in uranium-based hydrides. In $UH_3$ and related systems, U–H bonding strongly depopulates the U 6$d$ and 7$s$ conduction states, leaving electrical transport dominated by heavy U 5$f$ states with reduced mobility and enhanced sensitivity to disorder [7] [30].

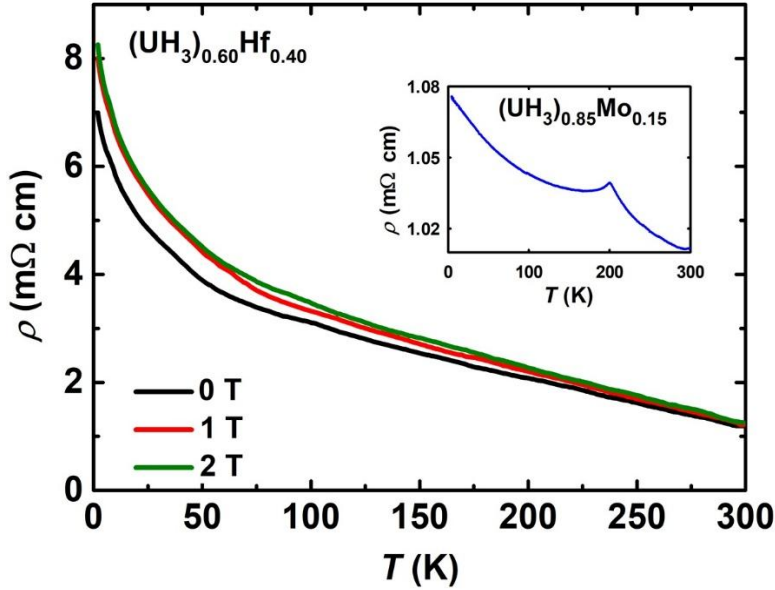

Fig. 10. Temperature dependence of electrical resistivity $\rho(T)$ for $(UH_3)_{0.60}Hf_{0.40}$ hydride measured in zero field and in applied fields 1 T and 2T, all measured in cooling cycle. For comparison, the inset shows the resistivity plot for $(UH_3)_{0.85}Mo_{0.15}$ hydride.

As a result, high resistivity and deviations from conventional metallic behavior are intrinsic features of this material class. In the present compound, the large Hf substitution further perturbs the local uranium environment relevant for transport, pushing the system deeper into a strong-disorder regime than more lightly substituted $UH_3$-based hydrides. This picture is further corroborated by the systematic reduction of the Sommerfeld coefficient γ with increasing Hf content in $(UH_3)_{1-x}Hf_x$ hydrides, indicating a reduced effective low-energy electronic density of states associated with U-derived states. Such a reduction signifies a diminished coherent 5$f$ contribution to charge transport rather than improved metallicity, thereby reinforcing incoherent transport and enhancing the relative importance of disorder and magnetic scattering, consistent with the robust negative resistivity slope observed in $(UH_3)_{0.60}Hf_{0.40}$.

Comparison with other $UH_3$-based hydrides highlights the specific features of the Hf-substituted system. Mo- and Ti-substituted $UH_3$ hydrides, such as $(UH_3)_{0.85}Mo_{0.15}$ (see inset of **Fig**.10) and $(UH_3)_{0.78}Mo_{0.12}Ti_{0.10}$, also exhibit large resistivities and negative slope [12] [31]. In those compounds, however, a cusp in $\rho(T)$ is typically observed at the Curie temperature, and the negative slope in the ordered state can be partially suppressed by an applied magnetic field due to field-induced reduction of static magnetic disorder [32]. The absence of a comparable cusp and the weak field response in $(UH_3)_{0.60}Hf_{0.40}$ indicate that the relevant disorder is more robust and less field-dependent. These observations suggest a transport mechanism dominated by disorder effects, with a significant contribution from the uranium magnetic subsystem. Uranium 5$f$ states are characterized by strong spin–orbit coupling and large orbital moments, leading to pronounced magnetic anisotropy and strong coupling between magnetism and charge transport [33] [34]. Consequently, magnetic disorder acts as an efficient and temperature-dependent scattering mechanism. In Mo- and Ti-substituted $UH_3$ hydrides, the field sensitivity of $\rho(T)$ below the Curie temperature provides direct evidence for the role of static magnetic disorder [32]. In contrast, the persistence of the negative resistivity slope under applied field in $(UH_3)_{0.60}Hf_{0.40}$ suggests that magnetic disorder and its coupling to charge carriers are not readily reduced by moderate magnetic fields, leaving temperature as the primary tuning parameter.

Within this framework, the negative resistivity slope in $(UH_3)_{0.60}Hf_{0.40}$ is best understood as a consequence of transport in an extremely strong-scattering regime, where heavy-carrier conduction is promoted by U–H bonding, strong disorder, and the temperature evolution of a disordered uranium magnetic subsystem that acts cooperatively. The smoothness of $\rho(T)$ and the absence of a clear ordering anomaly suggest a continuity between dynamic magnetic disorder at higher temperatures and static or quasi-static disorder at low temperatures, rather than a sharp transport-distinct transition. Finally, the data provide no evidence for a Kondo mechanism: there is no resistivity minimum, no logarithmic temperature dependence, and no field-induced suppression of the negative resistivity slope, consistent with previous studies of uranium-based hydrides and actinide intermetallics [32] [35].

## IV CONCLUSIONS

We demonstrated Hf substitution as an effective route to stabilize the metastable $\alpha$-$UH_3$ phase in binary uranium hydrides. Hydrogenation of *bcc*-$\gamma$ U-Hf alloys yields $(UH_3)_{1-x}Hf_x$ hydrides with a clear evolution phase content. Power XRD confirms that increasing Hf concentration progressively suppresses the stable $\beta$-$UH_3$ phase in $(UH_3)_{1-x}Hf_x$ hydrides. The $\alpha$-$UH_3$ dominates at $x = 0.30$ and the $\beta$-$UH_3$ phase is nearly fully suppressed at $x = 0.40$, analogous to the stabilization reported previously in Zr-alloyed $UH_3$. All $(UH_3)_{1-x}Hf_x$ hydrides are ferromagnetic with Curie temperatures in the range $T_C \approx 178\text{-}185$ K. The highest $T_C$ occurs near $x = 0.15$, after which it decreases slightly at higher Hf concentrations. Determination of $T_C$ by various methods (low-field M(T) data, Arrott plots, AC-susceptibility, and $C_p$(T)) shows an increasing uncertainty at larger $x$, consistent with a broadened magnetic transition. Concomitantly, the spontaneous magnetization is strongly reduced with Hf substitution, down to 0.46 $\mu_B$/U for $(UH_3)_{0.60}Hf_{0.40}$ from 1.0 $\mu_B$/U for $\beta$-$UH_3$, which is markedly different from the Zr-stabilized $\alpha$-$UH_3$ system, where the magnetization remains nearly unchanged over comparable substitution levels. The magnetic response is also strongly broadened at higher Hf concentrations as the inverse susceptibility of $(UH_3)_{0.60}Hf_{0.40}$ shows a pronounced low-field downturn above $T_C$ that is suppressed in high fields, consistent with ferromagnetic inhomogeneity (Griffiths-like behavior) expected for dilution/disorder in the itinerant 5$f$ ferromagnet. The low-temperature hysteresis loops are very broad and do not reach true saturation even at 14 T. The erratic remagnetization jumps at 2 K exhibited in $(UH_3)_{1-x}Hf_x$ hydrides resemble those in Zr-alloyed $UH_3$ [9], indicating strong pinning and disorder-controlled domain dynamics.

Specific-heat measurements further quantify how Hf modifies low-energy electronic and magnetic contributions in the stabilized hydrides. The Curie anomaly evolves from a sharper feature in the mixed phase $(UH_3)_{0.85}Hf_{0.15}$ to a broad anomaly in the $\alpha$-$UH_3$ rich compositions, closely resembling $(UH_3)_{0.70}Zr_{0.30}$. In contrast, some Zr compositions (e.g., $(UH_3)_{0.89}Zr_{0.11}$) show a well-defined cusp at $T_C$. Sommerfeld coefficient $\gamma$ decreases monotonically from 26.3 to 20.8 mJ/mol K$^2$, indicating a systematic suppression of the low-energy electronic term with increasing Hf content. Since $\gamma \propto n(E_F)$, the decrease in $\gamma$ indicates a reduction of the effective electronic density of states at the Fermi level with increasing Hf content, consistent with weakened low-energy 5$f$-derived electronic/magnetic response and, consequently, a smaller ordered moment. This trend correlates with the decrease in spontaneous magnetization value in $(UH_3)_{0.60}Hf_{0.40}$. The recovered magnetic entropy decreases systematically with Hf content (**TABLE** II) and reaches $S_{mag} \approx 3.1$ J/mol K $\approx 0.51$ R ln2 for $(UH_3)_{0.60}Hf_{0.40}$. Together with the field-sensitive non-Curie-Weiss susceptibility above $T_C$, the reduced $S_{mag}$ supports a picture of diluted or disordered 5f ferromagnetism, in which a substantial part of the magnetic entropy is released over a broad temperature interval rather than in a sharp transition.

Transport in $\alpha$-$UH_3$ stabilized $(UH_3)_{0.60}Hf_{0.40}$ hydride is highly anomalous, as $\rho$(T) is very large (m$\Omega$ cm) and shows a robust negative temperature coefficient from 2-300 K, lacks a clear cusp at its $T_C$, and exhibits only a weak positive magnetoresistance. The magnitude of $\rho$(T) places the system deep in a strongly incoherent, disorder-dominated regime (Mooji-correlation phenomenology), and comparison with Mo/Ti-substituted hydrides, where a cusp at $T_C$ and stronger field suppression are typically observed, highlights the more robust, large field-insensitive disorder in the Hf-substituted case.

## ACKNOWLEDGEMENT

The work was supported by the Grant Agency of Charles University (GA UK) under project no. 138124 and also supported by the Czech Science Foundation under the grant No. 25-16339S. MGML (http://mgml.eu/), supported within the Czech Research Infrastructure program (project no. LM2023065), is acknowledged for access to its technological and experimental facilities.